\documentclass[12pt]{article}

\usepackage{mathptmx,amsmath,amssymb,amsfonts}
\usepackage{graphicx,color,pictex,epstopdf,url,algorithm,algorithmic,caption}
\usepackage{etoolbox} % another kludge to over-ride Latex defaults
\makeatletter
\patchcmd{\@makechapterhead}{\large}{\normalsize}{}{}% for \chapter
\patchcmd{\@makechapterhead}{\large}{\normalsize}{}{}% for \chapter
\patchcmd{\@makeschapterhead}{\normalsize}{\normalsize}{}{}% for \chapter*
\makeatother
\usepackage{titlesec}
\titleformat*{\section}{\bfseries}
\titleformat*{\subsection}{\normalsize\bfseries}
\titlespacing\section{0pt}{12pt plus 3pt minus 2pt}{3pt plus 1pt minus 1pt}

\usepackage[textwidth=6.5in,textheight=9.1in,left=0.75in,right=0.75in,top=0.75in,bottom=0.75in]{geometry}
\makeatletter
\g@addto@macro\normalsize{\setlength\abovedisplayskip{3pt}}
\g@addto@macro\normalsize{\setlength\belowdisplayskip{3pt}}
\makeatother
\newtheorem{theorem}{Theorem}[section]
\newtheorem{lemma}{Lemma}[section]
\newtheorem{corollary}{Corollary}[section]

\newtheorem{remark}{Remark}[section]
\usepackage{parskip}
\let\oldref\ref
\renewcommand{\ref}[1]{(\oldref{#1})}  % stupid kludge for laTeX
\font\titlefonrm=ptmb scaled \magstep3
\font\figurefont=ptmr
\font\sc=cmcsc10
\newcommand{\re}{\mathop{\rm Re} }

\newcommand{\N}{\mathbf{N}}

\newbox\boxaddrone \newbox\boxaddrtwo

\title{\titlefonrm Recovery of a potential in a fractional diffusion equation}
\author{William Rundell\footnote{
Department of Mathematics,
Texas A{\small \&}M University, College Station, Texas 77843 USA.
rundell@math.tamu.edu}
\and
Masahiro Yamamoto\footnote{
Department of Mathematical Sciences,
The University of Tokyo, Tokyo 153 Japan.
myama@ms.u-tokyo.ac.jp
}
}
\begin{document}
%\date{}
\maketitle
\abstract{
We consider the determination of an unknown potential $q(x)$
form a fractional diffusion equation subject to overposed lateral
boundary data.
We show that this data allows recovery of two spectral sequences
for the associated inverse Sturm-Liouville problem and these are sufficient
to apply standard uniqueness results for this case.

We also look at reconstruction methods and in particular examine the issue
of stability of the solution with respect to the data.
The outcome shows the inverse problem to be severely ill-conditioned
and we consider the differences between the cases of fractional and of
classical diffusion.
}

%%%%%%%%%%%%%%%%%%%%%%%%%%%%%%%%%%%%%%%%%%%%%%%%%%%%

\section{Introduction}

A standard inverse problem is to recover a coefficient 
in an elliptic operator $\mathbb{L}$ from the diffusion equation
$u_t + \mathbb{L} u = 0$ from a combination of initial data $u_0(x)=u(x,0)$
and over-specified boundary data.
For example, with $\mathbb{L} =  -u_{xx} + q(x)u$ on the domain
$(0,1)\times (0,T)$ we might impose homogeneous boundary conditions,
say the flux, $u_x(0,t)=u_x(1,t)$ and measure the
data $u(0,t)=g(t)$ from which we would
hope to recover $q(x)$ for a given initial value $u_0(x)$.
Of course, we might also reverse the type of the boundary conditions.
Another possibility is to choose homogeneous initial conditions $u_0(x)=0$
and lateral conditions at $x=0$, $u(0,t)=0$, but now impose Cauchy data on
the other lateral boundary $x=1$.

Some of these approaches were taken in \cite{Pierce:1979} for
the case of the parabolic operator,
but we are also interested in the subdiffusion model involving
fractional derivatives and extending the parabolic case to one of a
subdiffusion process using a Djrbashian-Caputo fractional derivative
with index $\alpha$, $0<\alpha\leq 1$.
Fractional diffusion equations with Caputo derivatives in time have been widely
used as model equations for describing the anomalous diffusion phenomena.
Two important cases are highly heterogeneous aquifers and complex
viscoelastic material; see
\cite{BerkowitzCortisDentzScher:2006, HatanoHatano:1998} and also
\cite{SokolovKlafterBlumen:2002} for further applications.

As a by-product of the analysis we will expand upon known results for the
parabolic case $\alpha=1$.
The primary goal of this paper is to establish uniqueness results,
but we will also compare the degree of ill-conditioning of the problem
with respect to $\alpha$.
The latter is also of physical interest as it indicates
whether model reconstruction
problems for the fractional case differ substantially from the classical.
The now classical example of the backwards diffusion problem
for both the classical and fractional cases illustrates this possibility
where the degree of ill-conditioning differs remarkably
\cite{LiuYamamoto:2010, SakamotoYamamoto:2011}.
However, depending on the value of the final time $T$ this may not
translate into a superior numerical recovery for the backwards fractional
case $\alpha<1$, as shown in \cite{JinRundell:2015,KaltenbacherRundell:2019}.

\medskip

Specifically, we shall consider the following problem;
suppose $u(x,t)$ satisfies
\begin{equation}\label{eqn:direct_prob}
\begin{aligned}
&{}^C_0 D^\alpha_t u(x,t) - u_{xx}(x,t) + q(x) u(x,t) = 0,\qquad 0<x<1,\quad t>0 \\
&u_x(0,t) = 0,\quad u_x(1,t) = a(t),\qquad t>0 \\
&u(x,0) = 0,\quad 0\leq x\leq 1. \\
\end{aligned}
\end{equation}
Here ${}^C_0 D^\alpha_t$ denotes the Djrbashian-Caputo
fractional derivative of order $\alpha$, $0<\alpha\leq 1$,
with starting point the left-hand boundary $x=0$.
The potential $q(x)$ is assumed to be unknown and in order to utilize existing
results for \ref{eqn:direct_prob}, we take $q(x)\in  L^\infty$
although weaker conditions, for example $q\in L^2$ would suffice if
we only consider the  question of uniqueness.
We also might impose a nontrivial value of $u_0(x) = u(x,0)$.
However, regularity of the direct problem,
namely given $q(x)$ determine $u(x,t)$,
becomes a delicate issue in terms of the smoothness imposed on $u_0(x)$
and we prefer to avoid issues that are tangential to the main theme.
We remark that in general the fractional order operator in
\ref{eqn:direct_prob} has limited smoothing properties and this, together with
the nonhomogeneous version, \ref{eqn:direct_prob2} to be considered below,
is such that the solution $u$ has regularity that
depends strongly on the initial data, \cite{SakamotoYamamoto:2011}.

The current work also has ideas in common with \cite{CNYY:2009} where
the unknown coefficient appeared in the operator as a diffusion coefficient
$a(x)$, $\mathbb{L}u := (a(x)u_x)$ although in this work the boundary
conditions were homogeneous and the initial data was $u_0(x)=\delta(x)$.
If in the current situation we had instead the operator taken this operator
then  the inverse Sturm-Liouville uniqueness will still go through
but the analysis of reconstruction would require modifications.

Under the above conditions, there is a unique solution to \ref{eqn:direct_prob}
for any sufficiently smooth $a(t)$ and any $\alpha$, $0<\alpha\leq 1$,
see \cite{SakamotoYamamoto:2011}.

For reasons that will become apparent we shall restrict $a(t)$ to be integrable
and  have compact support on the interval $[0,T]$ for some fixed $T>0$.

Our goal is in addition to measure the flux data
\begin{equation}\label{eqn:overposed_data}
u_x(1,t) = b(t),\qquad t>T 
\end{equation}
and from the pair $\{a(t),b(t)\}$ seek to determine the unknown potential
$q(x)$.
\bigskip

\section{Background for fractional operators}

An essential component of fractional derivative problems is
the two-parameter Mittag-Leffler function $E_{\alpha,\beta}(z)$ defined by
\begin{equation}\label{eqn:mlf}
  E_{\alpha,\beta}(z) = \sum_{k=0}^\infty \frac{z^k}{\Gamma(\alpha k+\beta)}\quad z\in \mathbb{C},
\end{equation}
for $\alpha>0$, and $\beta\in\mathbb{R}$.
This generalizes the exponential function ubiquitous to classical diffusion;
$E_{1,1}(z) = e^z$.

\begin{lemma}\label{lem:mlf-entire}
For any $\alpha>0$ and $\beta\in\mathbb{R}$, $E_{\alpha,\beta}(z)$
is an entire function of order $\frac{1}{\alpha}$ and type 1.
\end{lemma}

\begin{lemma}\label{lem:ML_recurrence}
For $0< \alpha\leq 1$ and $x>0$, $\lambda>0$
\begin{equation}\label{eqn:ML_recur1}
\alpha\, \lambda \frac{d\ }{dx} E_{\alpha,1}(-\lambda x) = -E_{\alpha,\alpha}(-\lambda x)
\end{equation}
For $\re(\alpha)>0$ and $\re(\beta)>1$ and from $\lambda$ real
\begin{equation}\label{eqn:ML_recur2}
\frac{d\ }{dx}\, x^{\beta-1} E_{\alpha,\beta}(\lambda x^\alpha)
= x^{\beta-2} E_{\alpha,\beta-1}(\lambda x^\alpha)
\end{equation}
For $\Re(\alpha)>0$ and $\Re(\beta)>0$ and $a$ real
\begin{equation}\label{eqn:ML_recur3}
\frac{d\ }{dz} E_{\alpha,\beta}(a z) =
\frac{a}{\alpha z}\bigl(
E_{\alpha,\beta-1}(a z) - (\beta-1)E_{\alpha,\beta}(a z) \bigr)
\end{equation}
\end{lemma}

\begin{lemma}\label{lem:mlf-asymptotic}
Let $\alpha\in(0,1]$, $\beta\in\mathbb{R}$, $z\geq 0$,
and $N\in\mathbb{N}$.
Then with $z\to\infty$,
\begin{equation}\label{eqn:mlf-asymp}
E_{\alpha,\beta}(-z) = \sum_{k=1}^N\frac{(-1)^{k-1}}{\Gamma(\beta-\alpha k)}
\frac{1}{z^k} + O\Bigl(\frac{1}{z^{N+1}}\Bigr).
\end{equation}
\end{lemma}

\medskip
Following standard practice we transform equations~\ref{eqn:direct_prob}
into a set with homogeneous boundary conditions using
$v(x,t) = u(x,t) - a(t)$  to obtain
\begin{equation}\label{eqn:direct_prob2}
\begin{aligned}
&D^\alpha_t v(x,t) - v_{xx}(x,t) + q(x) v(x,t) = f(x,t),
\qquad 0<x<1,\quad t>0 \\
&v_x(0,t) = 0,\quad v_x(1,t) = 0,\qquad t>0 \\
&v(x,0) = 0,\quad 0\leq x\leq 1 \\
\end{aligned}
\end{equation}
with $f(x,t) = -(D^\alpha_t a(t) + q(x) a(t))$.

We assume that $D^\alpha_t a(t) \in L^\infty(0,T)$.
With $\Omega=(0,1)$, then $f\in L^2(\Omega\times(0,t^*))$ for any fixed $t^*$
and there exists a unique weak solution
$v(x,t)\in L^2(0,t^*; H^2(\Omega)\cap H^1_0(\Omega))$ such that
\begin{equation}\label{eqn:regularity}
\|u\|_{L^2(0,t^*),H^2(\Omega)} + D^\alpha_t u\|_{L^2(0,t^*),L^2(\Omega)}
\leq C\|f\|_{L^2(\Omega)\times(0,t^*))}
\end{equation}
See \cite[Theorem~2.2]{SakamotoYamamoto:2011}.

The solution to \ref{eqn:direct_prob2} is easily obtained
by separation of variables.
Let $\{\lambda_j,\phi_j(x;q,\lambda_j)\}_1^\infty$ be the Neumann
eigenvalues and eigenfunctions of $-\phi_j'' + q\phi_j = \lambda_j \phi_j$,
that is, with $\phi'_j(0)=\phi_j(1)=0$.
Let $\{\tilde\phi(x)\}$ denote the eigenfunctions with the normalization
$\|\phi\|_{_L^2}=1$.
Then from \cite{JinRundell:2015,SakamotoYamamoto:2011} the solution to \ref{eqn:direct_prob2} is given by
\begin{equation}\label{eqn:forcing_represent}
v(x,t) = \sum_{j=1}^\infty 
\int_0^t (t-\tau)^{\alpha-1}E_{\alpha,\alpha}(-\lambda_j(t-\tau)^\alpha)
\langle f(\cdot,\tau),\,\tilde\phi_j\rangle\,d\tau\, \tilde\phi_j(x).
\end{equation}

However, it will be more convenient for our purposes to assume
an endpoint normalization, more typical of Sturm-Liouville theory and
therefore we will instead use the normalization
$\phi(1)=1$
in place of $\|\phi\|_{_L^2}=1$.
Under this assumption  we then set $\rho_j = \|\phi\|_{_L^2}$.
Also, without loss of generality, we may assume
$\lambda_j > 0$, for $j\in \N$. 

From \ref{eqn:forcing_represent} we now obtain for the original dependent
variable evaluated at the right-hand boundary
\begin{equation}\label{eqn:u1t_represent}
u(1,t) = \sum_{j=1}^{\infty}
 \rho_j \int^t_0 s^{\alpha-1}E_{\alpha,\alpha}(-\lambda_j s^\alpha) a(t-s)\,ds,
\quad 0 < t < T. 
\end{equation}
An integration by parts using the assumption  $a(0) = 0$ yields
\begin{equation}\label{eqn:eqn_m3}
\begin{aligned}
\int^t_0 s^{\alpha-1}E_{\alpha,\alpha}(-\lambda_j s^\alpha) a(t-s)\,ds
&= -\frac{1}{\lambda_j}\int^t_0 \frac{d}{ds}(E_{\alpha,1}(-\lambda_j s^\alpha)) a(t-s)\,ds \\
&= -\frac{1}{\lambda_j}a(t) - \frac{1}{\lambda_j}\int^t_0 E_{\alpha,1}(-\lambda s^\alpha) 
a'(t-s)\, ds,\\
\end{aligned}
\end{equation}
and so
\begin{equation}\label{eqn:eqn_m4}
u(1,t) = -\sum_{j=1}^{\infty} \frac{\rho_j}{\lambda_j}a(t)
- \sum_{j=1}^{\infty} \frac{\rho_j}{\lambda_j}
\int^t_0 E_{\alpha,1}(-\lambda_j s^\alpha)
a'(t-s)\,ds  \quad 0 < t < T. 
\end{equation}
We know that $\rho_n = c_0 + o(1)$ as $n\to \infty$,
\cite{RundellSacks:1992}, and so
$$ 
\sum_{j=1}^{\infty} \left\vert \frac{\rho_j}{\lambda_j}\right\vert < \infty.
$$
Therefore, setting
$$
b = -\sum_{j=1}^{\infty} \frac{\rho_j}{\lambda_j}, \quad
A(t) = \sum_{j=1}^{\infty} \frac{\rho_j}{\lambda_j} E_{\alpha,1}(-\lambda_jt^{\alpha}),
$$
we obtain $b < \infty$ and $A \in C[0,\infty)$.
We can rewrite \ref{eqn:eqn_m4} as
\begin{equation}\label{eqn:eqnm5}
u(1,t) = -ba(t) - \int^t_0 A(s)a'(t-s)ds
= \int^t_0 (-b-A(s))a'(t-s)\,ds, \quad 0 < t < T 
\end{equation}
since $a(0) = 0$, this implies $a(t) = \int^t_0 a'(t-s)\, ds$.
\vspace{0.2cm}

Now let
$\sigma_n = \Vert \psi_n\Vert^{-2}_{L^2(0,1)}$ and
set $c = -\sum_{j=1}^{\infty} \frac{\sigma_j}{\mu_j}$ and
$B(s) = \sum_{j=1}^{\infty} \frac{\sigma_j}{\mu_j}E_{\alpha,1}(-\mu_j s^\alpha)$.
Since $u(p)(1,t) = u(q)(1,t)$, $0 < t < T$, we have
$$
\int^t_0 (b+A(s))a'(t-s)\,ds = \int^t_0 (c+B(s))a'(t-s)\,ds, \quad
0 < t < T.
$$
Titchmarsh's theorem and the analyticity in $t$ of $A$ and $B$
for $\re t > 0$ yields
$$
b + A(t) = c + B(t), \quad 0 < t < \infty.
$$
Taking the  Laplace transform then implies
$$
\frac{b}{z} + \sum_{j=1}^{\infty} \frac{\rho_j}{\lambda}\frac{z^{\alpha-1}}{z^{\alpha}+\lambda}
= \frac{c}{z} + \sum_{j=1}^{\infty} \frac{\sigma_j}{\mu_j}\frac{z^{\alpha-1}}{z^{\alpha}
+\mu_j}
$$
for Re $z > 0$.  Multiplying with $z^{1-\alpha}$ and setting
$\eta = z^{\alpha}$, we obtain
\begin{equation}\label{eqn:poles-residues}
\frac{b}{\eta} + \sum_{j=1}^{\infty} \frac{\rho_j}{\lambda}\frac{1}{\eta+\lambda}
= \frac{c}{\eta} + \sum_{j=1}^{\infty} \frac{\sigma_j}{\mu_j}\frac{1}{\eta+\mu_j}
\end{equation}
for $\re \eta > 0$.
By analyticity with respect to $\eta$ we see that the two representations
in \ref{eqn:poles-residues} must agree and so both the pole locations
and their residues must be identical.
This gives
$$
b=c, \quad \lambda_j = \mu_j, \quad 
\rho_j = \sigma_j, \quad \mbox{for all}\ j \in \N.
$$

The Gel'fand-Levitan theory for the potential-form inverse Sturm-Liouville
problem will now yield the uniqueness result

\begin{theorem}\label{thm:uniquness}
Suppose that $a(t)$ has support on the interval $[0,T]$ and that
$D^\alpha_t a(t)\in L^\infty[0,T]$.
Then there is at most one solution $\{q(x),u(x,t)\}$ to
\ref{eqn:direct_prob} and \ref{eqn:overposed_data}.
\end{theorem}

Since we will need this construction for the computational examples
the proof of this fact and its relation to
\ref{eqn:direct_prob} will now be briefly presented.

\section{The inverse Sturm-Liouville problem}

We denote by $\phi(x;q,\lambda)$ the solution of
\begin{equation}\label{eqn:phi_def}
-\phi''(x) + q(x)\phi(x) = \lambda\phi(x),\qquad \phi(0)=0,\quad\phi'(0)=1.
\end{equation}
For each $q$ there is clearly a unique solution $\phi(x)$ to
\ref{eqn:phi_def}.
We will impose boundary conditions at $x=1$ and look for the associated
eigenvalue/eigenvector pairs $\{\lambda_n,\,\phi_n(x)\}_{n-1}^\infty$ and
so we should view the condition  $\phi'(0)=1$ as being a normalization
of the eigenfunctions.

\begin{lemma}\label{lem:isl_two_spec}
Let $q_1$ and $q_2\in L^2(0,1)$ be two potentials.
Suppose we are given that the Dirichlet eigenvalues  $\{\lambda_n\}$ of
\ref{eqn:phi_def} for each of $q_1$ and $q_2$ are identical; that is
$\phi_n(1;q_1,\lambda_n) = \phi_n(1;q_2,\lambda_n) = 0$.
If further, the eigenvalues $\{\mu_n\}$ for the case of Neumann boundary
conditions at $x=1$ are also identical; that is
$\phi'_n(1;q_1,\mu_n) = \phi'_n(1;q_2,\mu_n) = 0$, then $q_1=q_2$ a.e.
\end{lemma}

Lemma~\ref{lem:isl_two_spec} is the famous two spectrum result of Borg,
\cite{Borg:1946}.
Since the original paper there have been several proofs of this result
and we will outline one below since the underlying machinery will be needed
in a later section.

The Gel'fand-Levitan transformation maps solutions of \ref{eqn:phi_def}
with $q=q_1$ into solutions with $q=q_2$ and is given by
\begin{equation}\label{eqn:GL_transform}
\phi(x;q_2,\lambda) = \phi(x;q_2,\lambda) +\int_0^x K(x,t)\phi(t;q_1,\lambda)\,dt
\end{equation}
where $K(x,t)$ is independent of $\lambda$ satisfies the hyperbolic equation
\begin{equation}\label{eqn:K_eqn}
\begin{aligned}
&K_{tt} - K_{xx} + (q_1(x) - q_2(t))K = 0 ,\qquad 0<t\leq x \leq 1\\
&K(x,\pm x) = \pm \frac{1}{2}\int_0^x [q_2(s)-q_1(s)]\,ds,\quad
K(x,0) = 0,\qquad 0\leq x \leq 1.\\
\end{aligned}
\end{equation}
For a proof of this computation see, \cite{RundellSacks:1992}
or the original paper, \cite{Gelfand-Levitan:1951}.

Now suppose $\phi_n(1;q_1,\lambda_n) = \phi_n(1;q_2,\lambda_n)$ for each
positive integer $n$.
Then from \ref{eqn:GL_transform} it follows that
$\int_0^1 K(1,t)\phi(t;q_1,\lambda_n)\,dt = 0$ and from the completeness
of the Dirichlet eigenfunctions that $K(1,t) = 0$.
If now $\phi'_n(1;q_1,\mu_n) = \phi'_n(1;q_2,\mu_n)$ then in a similar manner
we obtain $K_x(1,t) = 0$.
Under these conditions $K(x,t)$ satisfies a homogeneous hyperbolic equation
in the region $\{(x,t):\; 0\leq t\leq x \leq 1\}$ with zero Cauchy data on
the line $x=1$. 
It must therefore be identically zero in this region and hence also
on the boundary line $x=t$, that is, $K(x,x) = 0$.
From the second equation in \ref{eqn:K_eqn} we immediately obtain
$q_1=q_2$ a.e.
This proof was first shown in \cite{Suzuki:1985} and used as the basis
for solving other inverse Sturm-Liouville problems in a constructive manner
in \cite{RundellSacks:1992}.

Some of these other inverse spectral problems include replacing
the second spectrum by an endpoint condition on the derivative at $x=1$.
That is, we are given the Dirichlet eigenvalues $\{\lambda_n\}$ and
together with the values of the derivative of the associated eigenfunctions at
$x=1$, $\phi'(1;q,\lambda_n)$.
This is easily converted to the previous case.
The common Dirichlet spectrum gives $K(1,t)=0$ as before while the condition
$\phi'(1;q_1,\lambda_n)=\phi'(1;q_2,\lambda_n)$
when used in \ref{eqn:GL_transform} immediately shows that 
$K_x(1,t)=0$.

The original Gel'fand-Levitan paper showed uniqueness when
the Dirichlet spectrum $\{\lambda_n\}$ was given together with the norming
constants $\rho_n:=\|\phi(x;q,\lambda_n)\|^2$.
With the above formulation we can easily convert 
endpoint problem data to norming constant data
as follows (see \cite{RundellSacks:1992}).

We can view equation~\ref{eqn:GL_transform} as mapping solutions of
equation~\ref{eqn:phi_def} with the zero potential onto that with potential $q$
through
\begin{equation}\label{eqn:GL_transform_0}
\phi(x;q,\lambda) = \phi(x;0,\lambda) +\int_0^x K(x,t)\phi(t;0,\lambda)\,dt
= \frac{1}{\sqrt{\lambda}}\bigl[
\sin(\sqrt{\lambda}) +\int_0^x K(x,t)\sin(\sqrt{\lambda})\,dt\bigr]
\end{equation}
Then if we differentiate the equation $-y'' +qy=\lambda y$ with respect to
$\lambda$ we obtain
$-\dot y'' +q\dot y = \lambda\dot y+y$
where $\dot y$ denotes $\frac{\partial y}{\partial \lambda}$.
Multiplying this by $y$, the original equation by $\dot y$
and subtracting gives
$y^2 = y''\dot y - \dot y''y$.
Integrating between $x=0$ and $x=1$ and setting $\lambda=\lambda_n$
(so $y$ becomes $\phi_n(x)$) we get
$$ \int^1_0 \phi^2_ndx = \dot\phi_n(1) \phi'_n(1)$$
and therefore
\begin{equation}\label{eqn:rh_data}
\rho_n = \dot \phi_n(1) \phi'_n(1) \quad \hbox{or}\quad
\phi'_n(1) = \frac{\rho_n}{\dot\phi_n(1)}.
\end{equation}
We ant to convert the data $\{\rho_n\}$ into end-point data
$\{\phi'_n(1)\}$ and so we need an expression for $\dot\phi_n(1)$.

If we differentiate \ref{eqn:GL_transform_0} in $\lambda$ we obtain
$$\dot\phi(x) = -\frac{1}{2\lambda^{3/2}} \phi(x) + \frac{1}{2\lambda}
\left\{\cos\sqrt{\lambda} + \int^1_0 tK(1,t) \cos\sqrt\lambda\,t\,dt\right\}.
$$
Since $\phi_n(1) = 0$ we get
\vskip-15pt
$$
\qquad \dot\phi_n(1) = \frac{1}{2\lambda} \left\{\cos\sqrt{\lambda_n} + \int^1_0
t K(1,t) \cos\sqrt{\lambda_n} \, t\, dt\right\}
$$
and so from \ref{eqn:rh_data} we obtain
\vskip-15pt
\begin{equation}\label{eqn:rho_endpt}
\qquad \phi'_n(1) = {\frac{2\lambda_n\rho_n}
{\cos \sqrt{\lambda_n} + {\int^1_0} tK(1,t) \cos\sqrt{\lambda_n}\, t\, dt}}.
\end{equation}
The Dirichlet spectrum $\{\lambda_n\}$ gives $K(1,t)$ as before and 
in \ref{eqn:rho_endpt} we immediately obtain $\rho_n$ from $\phi'_n(1)$.

We summarize this as follows,

\begin{lemma}\label{lem:isl_nc_endpt}
Suppose we are given the Dirichlet spectra $\{\lambda_n\}_1^\infty$
for a potential $q$ and in addition, one of
\begin{enumerate}
\item{} For each spectral value $\lambda_n$ we are given the endpoint
derivative $\phi'(1.;q,\lambda_n)$
\item{} For each spectral value $\lambda_n$ the $L^2$ norm of the
eigenfunction, $\rho_n = \|\phi(1.;q,\lambda_n)\|$.
\end{enumerate}
Then either $\{\lambda_n,\phi'(1.;q,\lambda_n)\}$ or 
$\{\lambda_n,\rho_n\}$ uniquely determines $q$.
\end{lemma}

From the representation \ref{eqn:rho_endpt} we immediately obtain that
\begin{corollary}\label{cor:isl_nc_endpt}
If we have the Dirichlet spectra $\{\lambda_n\}_1^\infty$ and in addition
the combination $[\phi'(1)]^2/\rho_n$ for each $n\geq 1$, then this determines
$q$ uniquely.
\end{corollary}

\begin{remark}
We can also replace \ref{eqn:overposed_data} by one measuring the flux
on the leftmost boundary by using an almost identical analysis.
\end{remark}

\smallskip
While aesthetically pleasing, the above analytic continuation-based proof
should indicate the likelihood of the problem being severely ill-conditioned.
The two spectrum version of the inverse Sturm-Liouville problem is only mildly
ill-conditioned (although we will have some caveats to add to this later)
and the problem is transitioning the data function $b(t)$ into the
precise location of the zeros and poles of its complex-valued Laplace transform.
In the parabolic case we must locate the zeros and poles located on the
negative real axis in $s$-space from values obtained by integrating the data
$b(t)$ against an exponentially decaying function to obtain $\hat b(s)$
for all $s>0$.
The fractional case modifies this by in essence replacing the variable
$s$ by $s^\alpha$ and indicates that it might to some degree and under
certain circumstances shorten 
the distance the data has to be analytically continued in order to recover
the zeros and poles of $\hat b$.
In order to explore this further we will look at a slightly different version
of the above uniqueness result that will involve the solution representations 
of the previous section and provide more insight.

The challenge is to recover both $\{\lambda_j\}$ and $\{\rho_j\}$
uniquely from the representations such as \ref{eqn:eqn_m4} and.
\ref{eqn:poles-residues}.
Once this has been achieved then Corollary~\ref{cor:isl_nc_endpt} shows
that there is a unique $q$ satisfying 
\ref{eqn:direct_prob} with
\ref{eqn:overposed_data} for $t>T$.
In addition, as we saw in the previous section,
there is a well-proven reconstruction algorithm
for recovering $q(x)$ from the spectral data.

One cannot expect the recovery of the $\{\lambda_j\gamma_j\}$
to be well-posed and the case $\alpha=1$ illustrates the difficulties.
Now $E_{1,1}(z) = e^z$ and so in the parabolic case
of $\alpha=1$ becomes
\begin{equation}\label{eqn:lambda_residue_parabolic}
b(t) = \sum_{j=1}^\infty \gamma_j \int_0^t
e^{-\lambda_j(t-\tau)}a(\tau) \,d\tau
= \sum_{j=1}^\infty \beta_j e^{-\lambda_j t}
\qquad \mbox{for} \ t>T.
\end{equation}
where $\beta_j = \gamma_j\int_0^T e^{\lambda_j \tau} a(\tau)\,d\tau$.
Equation~\ref{eqn:lambda_residue_parabolic} is a Dirichlet series from 
which the coefficients $\{\beta_j,\lambda_j\}$ can be uniquely determined.
This can  be seen by taking Laplace transforms; the values of $\lambda_j$
are identified as the locations of the poles of $\hat b$ and $\beta_j$
as the residues at these poles.
From this, in theory, $\phi_j'(1)$ can be found from $\beta_j$
which be recovered once $\lambda_j$ is determined.
Solving the Dirichlet series for its component terms is a notoriously
ill-posed problem (as it should since it is tantamount to analytic continuation).
In addition, while recovering $\phi_j$ from $\beta_j$ is mathematically
obvious once we have $\lambda_j$, the coupling constant is
$\int_0^T e^{\lambda_j \tau} a(\tau)\,d\tau$ which grows exponentially with $j$
(and with $T$)
so the computational feasibility is another matter entirely.
This will severely restrict both the maximum interval of support $[0,T]$
as well as the number of frequencies $\lambda_j$ that can be obtained.

On the other hand, when $\alpha<1$, Lemma~\ref{lem:mlf-asymptotic} shows that
the Mittag-Leffler function has
only polynomial growth for large, negative arguments and so we might expect
that the fractional diffusion case will be less severely conditioned than
the parabolic as in \cite{LiuYamamoto:2010},
and for this to be more evident the smaller the fractional exponent $\alpha$.
We shall investigate this in the next section.

\section{Reconstructing the spectral data}

%\begin{equation}\label{eqn:direct_prob2_sol}
%v(x,t) = \sum_{j=1}^\infty\int_0^t (t-\tau)^{\alpha-1}E_{\alpha,\alpha}(-\lambda_j(t-\tau)^\alpha)(f,\phi_j)\phi_j \,d\tau
%\end{equation}
We shall examine a few special cases for the data $a(t)$.

\bigskip
Take $a(t)=1$ in $(0,T)$.
Reverting back to $\alpha\leq1$ in the representation
\ref{eqn:lambda_residue_parabolic}
and Using \ref{eqn:ML_recur2} with $\beta=1+\alpha$ and $x=t-\tau$ in
gives
\begin{equation}\label{eqn:forward_map}
\begin{aligned}
b(t) &= -\sum_{j=1}^\infty \gamma_j \int_0^T \frac{d\ }{d\tau}
\bigl[(t-\tau)^\alpha E_{\alpha,\alpha+1}(-\lambda_j(t-\tau)^\alpha) \bigr]
\,d\tau\\
&= -\sum_{j=1}^\infty \gamma_j\,
 (t-\tau)^\alpha E_{\alpha,\alpha+1}(-\lambda_j(t-\tau)^\alpha)
\Big|_{\tau=0}^{\tau=T}
\end{aligned}
\end{equation}
Suppose the goal is to recover the first $N$ elements of the spectral
sequence pair $\{\gamma_j,\lambda_j\}_{j=1}^N$ from \ref{eqn:forward_map}.
Then we define $F:\mathbb{R}^{2N} \to C(T,\infty)$ by $F$ corresponding
to the first $N$ terms on the right hand side of \ref{eqn:forward_map}
\begin{equation}\label{eqn:F_map}
\begin{aligned}
F(\{\lambda_j,\phi_j'(1)\}_1^N) &= 
\sum_{j=1}^N \gamma_j\bigl[
t^\alpha E_{\alpha,\alpha+1}(-\lambda_j t^\alpha)
- (t-T)^\alpha E_{\alpha,\alpha+1}(-\lambda_j(t-T)^\alpha)
\bigr]\\
&=: \sum_{j=1}^N \gamma_j K_\alpha(t,\lambda_j)\\
\end{aligned}
\end{equation}
We then seek a solution of the nonlinear equation
\begin{equation}\label{eqn:F_map2}
F(\{\lambda_j,\phi_j'(1)\}_1^N) = b(t)
\end{equation}
for the eigenvalues and endpoint values.
Note that the range of $F$ is in fact analytic so that in
the values over any time interval suffices in theory to
determine the values for all complex $t$.
However, we are now interested in the question of a feasible reconstruction
of the spectral data and it may seem that choosing a large range of $t$
values will give a more accurate representation of the series
especially under a situation where the measured values of $b$ are subject to
uncertainty.

The function $K_\alpha(t,\lambda)$  has the same large $t$
asymptotic behaviour for all $\alpha$; from \ref{eqn:mlf-asymp} we see that
$E_{\alpha,\alpha+1}(-z) = \frac{1}{\Gamma(1)}\frac{1}{z} + O(\frac{1}{z^2})$
for $z>0$. 
Thus a little algebra shows the kernel $K_\alpha$ can be expected to decay
as $O\bigl(t^{-2}\bigr)$ for $t>\!> T$ and any $\alpha<1$.
That is, the asymptotic decay of $K_\alpha$ is, up to a constant
multiplier, independent of $\alpha$
provided $\alpha<1$.
This is in sharp contrast to when $\alpha=1$ and
shows that taking measurements for large times as a means
of recovering eigenvalues beyond the first few is pointless in the classical
heat equation as the value of the kernel becomes exponentially small.
On the other hand, in the fractional diffusion case such large times are not
specifically excluded on this count but there is a difficulty for small values of $t$. 
While the Mittag-Leffler functions decay of polynomial order for large,
negative argument, for small time values, due to the fractional power
$t^\alpha$ the values of $K_\alpha(t,\lambda)$ for $\alpha<1$ are be less than
those for $\alpha=1$ indicating an advantage to the parabolic case over
this range.
This is precisely the effect found in the backwards diffusion
problem discussed in \cite{JinRundell:2015} and the unknown source location
problem from time-data in\cite{RundellZhang:2017}.

However, none of this gives insight into the actual inversion of
\ref{eqn:F_map} which would require looking at the derivative of 
$F$ with respect to the parameters $\{\lambda_j,\gamma_j\}$.

If we now take $a(t)=\delta_T(t)$ then the previous constructions become
\begin{equation}\label{eqn:lambda_residue_delta}
b(\tilde t+T) = \sum_{j=1}^\infty \gamma_j 
\tilde t^{\alpha-1}E_{\alpha,\alpha}(-\lambda_j \tilde t^\alpha)
\qquad \mbox{for} \ \tilde t>0.
\end{equation}
In this case the kernel $K_\alpha$
now involves the function $E_{\alpha,\alpha}(-z)$.
This again has quadratic decay for large positive arguments $z$
the reason being that the term in $\frac{1}{z}$ of the asymptotic expansion
is missing since $1/\gamma(\beta-\alpha) \to 0$ as $\beta\to\alpha$.
However, when one takes into account that $z=t^\alpha$ and the additional
singular term $t^{\alpha-1}$ the overall asymptotic behaviour is
$\frac{1}{\Gamma(-\alpha)}t^{-1-\alpha} + O\bigl(t^{-1-2\alpha}\bigr)$
and now is no longer independent of $\alpha$ and the constant in the leading
term approaches zero as $\alpha\to 1$.
%Thus in terms of the long time behaviour we should expect quite similar
%overall behaviour between the two cases provided 
%This is what one should expect from physical considerations.

A natural way to solve \ref{eqn:F_map2} or the form
\ref{eqn:lambda_residue_parabolic} for more general $a(t)$,
is to use Newton's method.
Computation of the derivative map is possible from the representation 
\ref{eqn:forward_map}. We also have a reasonable starting approximation
since  we know the asymptotic behaviour of both spectral sequences
and the asymptotic values are obtained to a high degree of approximation
for even relatively small $N$ provided $q$ is smooth.
As we will see, unless we have definite prior information about $q$,
this is an assumption that will be forced on us due to the inevitable
ill-conditioning of the problem.
In fact, the above suggests that computing the derivative about the
approximation $q=0$ should give the essential features of the problem
and this simplification has been a fairly standard approach
for this type of situation,
\cite{RundellSacks:1992,JinRundell:2012b,JinRundell:2015}.

As we must expect, the greatest difficulty lies in the extraction
of the eigenvalues and so we will look at the $M\times N$ submatrix
$\frac{\partial F}{\partial\lambda}$ where we assume that $M$ $t$-values
have been given over a subset of $(T,\infty)$ and the values of $\gamma_j$
are held at their asymptotic value.
This computation requires evaluating the derivative of 
$E_{\alpha,\alpha+1}(-\lambda t^\alpha)$ with respect to $\lambda$.
In the case $a(t)=\delta_T(t)$ we would obtain
$E_{\alpha,\alpha}(-\lambda t^\alpha)$.
%\footnote{\color{blue}{Maybe do instead for $a(t) = \delta_T(t)$??}}
From \ref{eqn:ML_recur3} with $\beta=1+\alpha$ we obtain
\begin{equation}\label{eqn:lambda_residue_delta}
\frac{\partial\ }{\partial\lambda} E_{\alpha,\alpha+1}(-\lambda s^\alpha)
= \frac{s^\alpha}{\alpha\lambda}\bigl[
 \alpha E_{\alpha,\alpha+1}(-\lambda s^\alpha) 
-E_{\alpha,\alpha}(-\lambda s^\alpha).
\bigr]
\end{equation}

Suppose now we have obtained the sequences $\{\lambda_j,\gamma_j\}_1^N$
for some $N$ where $\gamma_j = \bigl[\phi_j'(1)/\|\phi_j\|\bigr]^2$.
Then it is quite straightforward to reconstruct a potential $q_N(x)$ from
this data. 
We make the ansatz that $\{\lambda_j,\gamma_j\}_{j=N+1}^\infty$
are given by our best estimate of these values for a fixed potential;
The latter can be taken to be for $q=0$; a better option is for
$q(x)=\bar q := \int_0^1 q(s)\,ds$ where $\bar q$ can be
estimated from $\bar q \approx \lambda_N - N^2 \pi^2$
This estimate will be reasonable for modest size $q$ and relatively small $N$
provided $q$ is smooth, but degenerates outside of these conditions,
see \cite{RundellSacks:1992}.

The reconstruction from spectral data can be viewed as only mildly
ill-conditioned amounting to effectively only a derivative loss,
or in terms of the spectral data, control in the finite dimensional
$H^1$ norm controls $q$ in $L^2(0,1)$, \cite{McLaughlin:1986,RundellSacks:1992}.
However from a reconstruction perspective this is not the complete story.
The asymptotic behaviour $\lambda_n = n^2\pi^2 + \bar q + c_n(q)$
where $c_n = O\bigl(n^{-k+2}\bigr)$ for $q\in C^k[0,1]$,
shows that the information term $c_n$ is very small in comparison with
the masking term $n^2\pi^2$.
For even a modestly smooth $q$, say $q\in C^2[0,1]$,
this ``signal to background''
ratio can be easily of order $10^6\,-\,10^7$ for $n=10$ and only an order
of magnitude more for $n=5$,
see \cite{RundellSacks:1992}.
While a smooth function may not require as many Fourier modes for a reasonable
reconstruction, it does show that errors made in computing $\{\lambda_n\}$
will be magnified considerably when applied to $\{c_n\}$ and it is this
sequence that holds the information on $q$.

For the above reason, the reconstruction method of first reducing
to an inverse spectral problem, then recovering $q$ from spectral data
is not optimal. 
The inversion of \ref{eqn:F_map2} to obtain $\{\lambda_j,\gamma_j\}$ is severely
ill-posed and a regularization step must be applied.
One can certainly use truncated {\sc svd} but this is a rather blunt tool
in this context.
Tikhonov regularization not only requires estimating the regularization
parameter but also requires penalizing in some norm.
For the case of the $\{\lambda_n\}$ recovery we can build in the masking term
$n^2\pi^2$ and a prior assumption about the decay of the coefficients $\{c_k\}$
based on an assumption about the smoothness of $q$.
This means solving not for the eigenvalues themselves but writing
$\lambda_j = j^2\pi^2 + c_j$ in the definition of F in \ref{eqn:F_map2}.
This is less straightforward than simply penalizing against what prior
information one has on $q$ directly.

There is another aspect; the representation \ref{eqn:F_map2} in any of its
forms is valid only for $t>T$.
This is true even for the case $\alpha=1$ as
\ref{eqn:lambda_residue_parabolic} shows.
This restriction is not essential for the uniqueness proof provided
we avoid negative eigenvalues and this can be done by assuming
a lower bound for $q$.
The next section will give a more direct reconstruction algorithm  and show that
there is a considerable advantage to measuring the flux $b(t)$ as early
as possible.

\section{Reconstructing the potential}

Let $u(x,t;q)$ be the solution to \ref{eqn:direct_prob}
for a given $q(x)\in L^2(0,1)$.
Then define the map $\mathcal{F}(q)$ by 
\begin{equation}\label{eqn:Fq}
\mathcal{F}(q) = -u_x(1,t;q),\qquad \mbox{for\ } \;t \in  I_t 
\end{equation}
and we must solve $\mathcal{F}(q) = b(t)$.
Here $I_t$ is the measurement interval over which we measure the flux 
$u_x(1,t)$.
This can be the interval $(T,T_f]$ for some fixed final time $T_f$
as stated originally, or $(0,T_f)$ as suggested at the end of the last
section.

Following the line from the previous section, we propose to solve this by 
Newton's method.
This requires a computation of $\frac{\partial\ }{\partial q}\mathcal{F}$
and it is easily seen that
$\frac{\partial\ }{\partial q}\mathcal{F}[q].\delta q$ is the solution of
\begin{equation}\label{eqn:Fp_ibvp}
\begin{aligned}
&D^\alpha_t v(x,t) - v_{xx}(x,t) + q(x) v(x,t) = -\delta q(x)u(x,t;q_n),
\qquad 0<x<1,\quad t>0 \\
&v(0,t) = 0,\quad v(1,t) = 0,\qquad t>0 \\
&v(x,0) = 0,\quad 0\leq x<\leq 1 \\
\end{aligned}
\end{equation}
Then from an initial approximation $q_0(x)$ we have the following
recursion scheme to define $q_n(x)$ 
\begin{equation}\label{eqn:Newt_iter}
\frac{\partial\ }{\partial q}\mathcal{F}[q].\delta q = b(t) - \mathcal{F}[q_n],
\qquad q_{n+1} = q_n + \delta q.
\end{equation}
We can also look at the special case of a ``frozen Newton Scheme''
where the derivative is held at a fixed value of $q(x)$, in particular
when $q=0$.
This leads to a formulation very close to that of the previous section.
It also allows us to analyze the derivative of $\mathcal{F}$ as a function of
$\alpha$ which is a primary goal.

To this end we assume that $q(x)$ can be represented by a set of basis
functions $\mathcal{B} = \{\psi_n(x)\}_1^N$ for suitably chosen $\psi_n(x)$,
We thus evaluate $\mathcal{F}[q=0]$ over a set of $M$ sample points on the
interval $I_t$ for a fixed function $a(t)$ with support on $[0,T]$ and
where the directions $\delta q$ are taken from $\mathcal{B}$.
Our goal is investigate the distribution of the singular values of the
corresponding matrix
$$
J_\alpha = \mathcal{F}[0].\psi_k(t_j), \qquad t_j \in I_t,\quad 1\leq k\leq N
$$
as the fractional derivative constant $\alpha$ takes on values in $(0,1]$.

We should make some comment on time scales.
In \ref{eqn:direct_prob} physical constants have been normalized to unity.
In particular, it is rescaled with a unit diffusion coefficient and a
more physically accurate version would incorporate a diffusion coefficient
$c(x)$ in the elliptic operator, that is,
$D^\alpha_t u(x,t) - c(x)u_{xx}(x,t) + q(x) u(x,t) = 0$.
This coefficient might itself be a ratio of conductivity and
specific heat and can vary considerably from material to material.
Typically it will be much smaller than unity 
(see for example the discussion in \cite[Section~3.1]{JinRundell:2015})
and this rescaling affects
the potential $q(x)$ and the spectrum of the elliptic operator.
If these are kept at the normalized values they in turn affect the time scales
under consideration -- and there will be an $\alpha$-dependence here.
We shall ignore this and choose to work with a unit coefficient
recognizing that from a physical perspective this leads to inflated
time scales.

\newdimen\xfiglen \newdimen\yfiglen
\xfiglen=2.7 true in
\yfiglen=1.6 true in
\newbox\figurelegend
\newbox\figureone
\newbox\figuretwo
\newbox\figurethree
\newbox\figurefour
\newbox\figurefive
%%%%%%%%%%%%%%%%%%

\setbox\figurelegend=\hbox{
\small
\beginpicture
  \setcoordinatesystem units <0.7true in,0.8true in> point at 0 0
  \setplotarea x from 0 to 1, y from 0 to 0.6
\footnotesize
  \put {{\color{red}$\diamond$}}\relax [l] at 0 0.6
  \put {$\alpha = 0.25$} [l] at 0.3 0.6
  \put {{\color{green}$\ast$}}\relax [l] at 0 0.4
  \put {$\alpha = 0.5$} [l] at 0.3 0.4
  \put {{\color{blue}$\circ$}}\relax [l] at 0 0.2
  \put {$\alpha = 0.75$} [l] at 0.3 0.2
  \put {{\color{black}$\bullet$}}\relax [l] at 0 0.0
  \put {$\alpha = 1$} [l] at 0.3 0.0
\endpicture
}
\setbox\figureone=\vbox{\hsize=\xfiglen
\beginpicture
\footnotesize
  \setcoordinatesystem units <0.1\xfiglen,0.1\yfiglen> point at 1 -10.5
  \setplotarea x from 1 to 10, y from -10 to 1.5
  \axis bottom shiftedto y=-10 ticks short numbered from 1 to 10 by 1 /
  %\axis left ticks short withvalues $10^{-10}$ $10^{-8}$ $10^{-6}$ $10^{-4}$ $10^{-2}$ $10^{0}$ / at -10 -8 -6 -4 -2 0 / /
  \axis left ticks short withvalues ${-10}$ ${-8}$ ${-6}$ ${-4}$ ${-2}$ ${0}$ / at -10 -8 -6 -4 -2 0 / /
\small
\put {\copy\figurelegend} [rt] at 10 0.5
\put {$k$} [lb] at 10.1 -10
\put {$\log_{10}(\sigma_k)$} [lb] at 1.1 1
%
% alpha = 0.25
\multiput {{\color{red}$\diamond$}} at
1   0.830
2  -0.666
3  -2.546
4  -4.680
5  -7.057 
6  -9.642
%7  -12.395
%8  -15.000
 /
%
% alpha = 0.5
\multiput {{\color{green}$\ast$}} at
1   0.827
2  -0.293 
3  -1.755
4  -3.428
5  -5.311
6  -7.360 
7  -9.478
%8  -11.517
%9  -13.481
%10 -15.000 
 /
%
% alpha = 0.75
\multiput {{\color{blue}$\circ$}} at
1   0.829
2  -0.036
3  -1.185 
4  -2.526
5  -4.050
6  -5.684
7  -7.262
8  -8.726 
9  -10.185 
%10 -11.871 
 /
%
% alpha = 1
\multiput {{\color{black}$\bullet$}} at
1   0.838
2   0.186
3  -0.680 
4  -1.737 
5  -2.952 
6  -4.258 
7  -5.479 
8  -6.627
9  -7.814
10 -9.248
 /
\endpicture
}
\setbox\figuretwo=\vbox{\hsize=\xfiglen
\small
\beginpicture
  \setcoordinatesystem units <0.1\xfiglen,0.1\yfiglen> point at 1 -10.5
  \setplotarea x from 1 to 10, y from -10 to 1.5
  \axis bottom shiftedto y=-10 ticks short numbered from 1 to 10 by 1 /
%  \axis left ticks short withvalues $10^{-10}$ $10^{-8}$ $10^{-6}$ $10^{-4}$ $10^{-2}$ $10^{0}$ / at -10 -8 -6 -4 -2 0 / /
  \axis left ticks short withvalues ${-10}$ ${-8}$ ${-6}$ ${-4}$ ${-2}$ ${0}$ / at -10 -8 -6 -4 -2 0 / /
\small
\put {\copy\figurelegend} [rt] at 10 0.5
\put {$k$} [lb] at 10.1 -10
%\put {$\sigma_k$} [lb] at 1.1 1
\put {$\log_{10}(\sigma_k)$} [lb] at 1.1 1
%
% alpha = 0.25
\multiput {{\color{red}$\diamond$}} at
1   0.6754
2  -0.8254
3  -2.7096
4  -4.8473
5  -7.2269
6  -9.8147
7  -12.5636
8  -13.2058
9  -14.5268
10 -14.7173
 /
%
% alpha = 0.5
\multiput {{\color{green}$\ast$}} at
1   0.6724
2  -0.4524
3  -1.9182
4  -3.5934
5  -5.4785
6  -7.5288
7  -9.6472
8  -11.6851
9  -13.3659
10 -13.7625
/
%
% alpha = 0.75
\multiput {{\color{blue}$\circ$}} at
1    0.6748
2   -0.1955
3   -1.3473
4   -2.6906
5   -4.2157
6   -5.8504
7   -7.4268
8   -8.8894
9  -10.3480
10 -12.0337
 /
%
% alpha = 1
\multiput {{\color{black}$\bullet$}} at
1   0.4461
2  -0.2097
3  -1.0777
4  -2.1354
5  -3.3501
6  -4.6523
7  -5.8615
8  -6.9931
9  -8.1545
10 -9.5536
 /
\endpicture
}
\setbox\figurethree=\vbox{\hsize=\xfiglen
\beginpicture
\footnotesize
  \setcoordinatesystem units <0.1\xfiglen,0.1\yfiglen> point at 1 -10.5
  \setplotarea x from 1 to 10, y from -10 to 1.5
  \axis bottom shiftedto y=-10 ticks short numbered from 1 to 10 by 1 /
%  \axis left ticks short withvalues $10^{-10}$ $10^{-8}$ $10^{-6}$ $10^{-4}$ $10^{-2}$ $10^{0}$ / at -10 -8 -6 -4 -2 0 / /
  \axis left ticks short withvalues ${-10}$ ${-8}$ ${-6}$ ${-4}$ ${-2}$ ${0}$ / at -10 -8 -6 -4 -2 0 / /
\small
\put {\copy\figurelegend} [rt] at 10 0.2
\put {$k$} [lb] at 10.1 -10
\put {$\log_{10}(\sigma_k)$} [lb] at 1.1 1
%\put {$\sigma_k$} [lb] at 1.1 1
%
% alpha = 0.25
\multiput {{\color{red}$\diamond$}} at
1   -0.4126
2   -1.9084
3   -3.7709
4   -5.8671
5   -8.1941
%6  -10.7522
%7  -13.4680
%8  -14.9041
%9  -16.2822
%10  -17.2721
 /
%
% alpha = 0.5
\multiput {{\color{green}$\ast$}} at
1   -0.4156
2   -1.5386
3   -2.9998
4   -4.6573
5   -6.5403
6   -8.6000
%7  -10.7731
%8  -13.4598
%9  -14.9340
%10  -16.1223
/
%
% alpha = 0.75
\multiput {{\color{blue}$\circ$}} at
1   -0.4132
2  -1.2831
3  -2.4361
4   -3.7683
5   -5.3248
6   -6.9544
7   -8.8667
%8 -11.4522
%9 -14.0827
%10-16.4465
 /
%
% alpha = 1
\multiput {{\color{black}$\bullet$}} at
1   -0.4038
2   -1.0595
3   -1.9288
4   -2.9804
5   -4.1988
6   -5.5915
7   -6.8596
8   -8.8276
%9  -11.7923
%10 -16.0493
 /
\endpicture
}

\begin{figure}[ht]
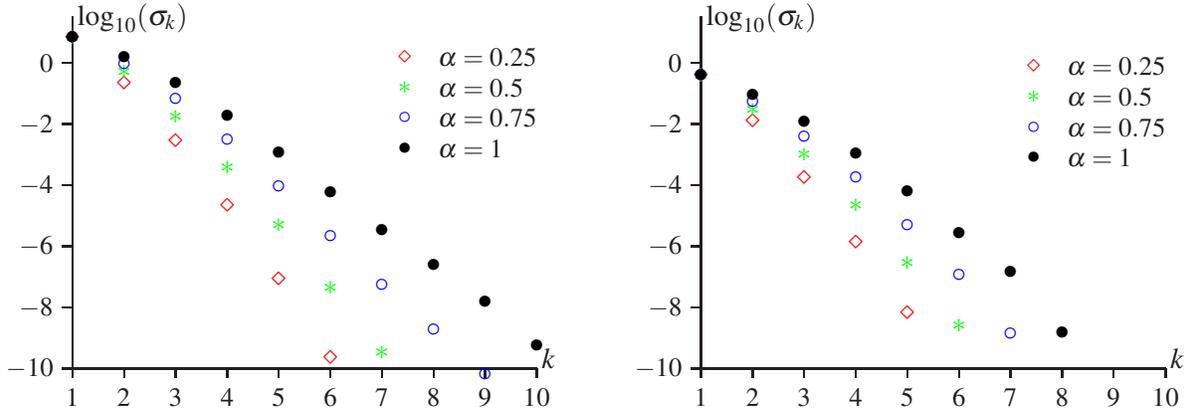

\centering
\captionsetup{justification=centering,margin=2cm}
\hbox to \textwidth{\hss\box\figureone\hss\box\figurethree\hss}
\small
\caption{\figurefont Singular values of $J_\alpha$ for
$\alpha =$ $\footnotesize{\frac{1}{4},\;\frac{1}{2},\,\frac{3}{4},\,1}$.
$\;T=1$, $I_t=(1,2]$.
Sampling within $I_t$ was at every $\delta t$;
the leftmost figure has $\delta t=0.001$, the rightmost $\delta t=0.05$.}
\label{fig:sv-Jac1}
\end{figure}

Figure~\ref{fig:sv-Jac1} shows the singular values of $J_\alpha$
for $\alpha = \frac{1}{4},\;\frac{1}{2},\,\frac{3}{4},\,1$ when
$T=1$ and $I_t=(1,2]$ where $a(t)$ is taken to be the function
$a(t) = \sin^3(\frac{\pi}{T}t)$.
Sampling within the measurement interval $I_t$ was taken at every
$\delta t$; the leftmost figure shows the case $\delta t=0.001$,
the rightmost the case $\delta t=0.05$.
Note that we are only seeking $10$ modes from the linearized map
so that both of these are oversampling, although the leftmost figure
exceeds this by a considerable amount.
This illustrates the extreme, likely exponential order,
ill-conditioning of the problem for all values of $\alpha$
and this increases with decreasing $\alpha$.

The explanation for this difference is as follows.
For the heat equation we are trying to extract the values of $\lambda_j$ from
$e^{-\lambda_j t_k}$. If $t_1$, the lowest sampled value, is large enough
so that  $e^{-\lambda_j t_1} < \delta$ where $\delta$ is a measure of our
measurement accuracy to handle small values then we will be unable
to recover this $\lambda_j$.
The more sample points taken, especially for small value of $t-T_0$
the better our recovery of, in particular, the larger eigenvalues.
Note also that the coefficient $\gamma_j$ will decrease with $\lambda_j$
adding to the effect.
In the case of $\alpha<1$, for small, negative values of its argument,
the Mittag-Leffler function initially decays
much faster than the exponential
(and this rate increases with decreasing $\alpha$)
-- again accentuating the phenomenon and providing a rationale for the figures.

If instead of measuring the flux $b(t)$ starting at $t=T$, that is,
immediately after the cut-off value of the support of $a(t)$, we delay
for an interval $T_1,T_2$ where $T_1>T$ then the picture changes.
The number of recoverable $\lambda_k$ decreases markedly and especially
for the parabolic case $\alpha=1$.
This is again what we should expect from the previous discussion.
The significant difference is now on the dependence of $\alpha$.
The rapid decay of the exponential function for even modest values of
$-\lambda t$ severely limits the utility of larger time measurements.
In the case of $\alpha<1$ the controlling Mittag-Leffler function decays
only polynomially for large negative argument and so large time measurements
remain useful.

As an example of the above, if we measure only over $[1.5T,3T]$ (with $T=1$)
then all singular values $\sigma_k$ for $k\geq 3$ are less than $10^{-10}$;
the first two singular values are approximately $10^{-2.2}$ and $10^{-4.3}$.
For $\alpha=\frac{1}{4},\;\frac{1}{2},\,\frac{3}{4}$ the first 3 singular
values are greater than $10^{-7}$ and the decay for the larger index singular
values becomes asymptotically nearly independent of $\alpha$ and significantly
greater than that for the case $\alpha=1$ as should be expected from the
asymptotic behaviour of the Mittag-Leffler function.
However, the magnitude of these singular values are still sufficiently small
to make the corresponding singular vectors unusable in almost any
practical application.
For $\alpha=\frac{3}{4}$ the first four singular values are approximately
$10^{-1.1}$, $10^{-3.3}$, $10^{-5.8}$, $10^{-7.5}$.
Thus if a rough approximation is sufficient this is possibly obtainable
in the fractional case, but unlikely in the classical.
On the other hand for an immediate measurement, especially with a high sampling rate, the opposite is true,

This reversal of the effective conditioning of the cases
$\alpha<1$ and $\alpha=1$ is similar to the situation with the backwards
diffusion problem noted in \cite{JinRundell:2015} -- although more complex.

In conclusion, one can see that while equations \ref{eqn:direct_prob} with
\ref{eqn:overposed_data} gives a unique potential $q(x)$ the inverse problem
is severely ill-posed.
This is yet another example of the ``folk theorem:'' that a problem where
the data is given in one direction (here time) and the unknown (here $q(x)$)
is given in an orthogonal direction is almost certain to be severely
ill-conditioned, \cite{Cannon:1975}.
Here the reason for this ill-conditioning comes in through each of the
reconstruction methods.
In the first, analytic continuation was used to obtain spectral
information on the operator ${\cal L}u := -u_{xx} + q u$, thereafter converting
the inversion into a mildly ill-conditioned one of known type.
In the second, a direct conversion method was used and the linearization of
the associated map formed.  The inversion of this map is equivalent to
a problem that is known to be severely ill-conditioned.

\section{Acknowledgment}
The work of the first author was supported in part by the
National Science Foundation through award DMS-1620138.
The work of the second author was supported by 
{\sc JSPS KAKENHI} Grant Number {\sc JP15H05740} and
by the A3 Foresight Program ‘Modeling and
Computation of Applied Inverse Problems’,
Japan Society for the Promotion of Science ({\sc JSPS}). 
%%%%%%%%%%%%%%%%%%%%%%%%%%%%%%%%%%%%%%%%%%%%
%\vfill\eject

\bibliographystyle{plain}
\bibliography{frac_pot.bib}

\end{document}